XeLatex=\RequirePackage{lineno}
\documentclass[aps, prl,
floatfix,twocolumn,superscriptaddress,
]{revtex4-2}
\usepackage{t1enc}
\usepackage[T1]{fontenc}
\usepackage{lmodern}
\usepackage{graphicx}
\usepackage{epsfig}
\usepackage{amssymb}
\usepackage{amsmath}
\usepackage{dcolumn}
\usepackage{bm}
\usepackage{color}
\usepackage{float}
\usepackage{hyperref}
\usepackage{natbib}
\usepackage{footnote}
\usepackage{ulem}
\usepackage{hhline}
\begin{document}

\title{Comment on: ``Crossover of Charge Fluctuations across the Strange Metal Phase Diagram''}

\author{J\"org\,Fink}
\affiliation{Leibniz Institute for Solid State and Materials Research  Dresden, Helmholtzstr. 20, D-01069 Dresden, Germany}
\affiliation {Institut f\"ur Festk\"orperphysik,  Technische Universität Dresden, D-01062 Dresden, Germany}

\date{\today}

\begin{abstract}\
In a recent paper by Husain {\it et al.} [PRX \textbf{9}, 041062 (2019)], the two-particle electronic excitations in Bi$_2$Sr$_2$CaCu$_2$O$_{8+x}$  have been studied by Electron Energy-Loss Spectroscopy in reflection (R-EELS) in the ``strange metal'' range between underdoped and overdoped materials. The authors conclude that there are no well defined plasmons. Rather they obtain a momentum-independent continuum which they discuss in terms of holographic theories. In this Comment it is pointed out that the experimental results are in stark contrast to previous EELS in transmission (T-EELS), Resonant Inelastic X-ray Scattering (RIXS), and  optical studies. The differences can be probably explained by an inaccurate momentum scale in the R-EELS experiments. Furthermore, it is shown, that many material specific experimental results from T-EELS, R-EELS, RIXS, and optical spectroscopy can be explained by a more traditional extended Lindhard model. This model describes the energy, the width, and the dispersion of normal and accustic plasmons in cuprates, as well as the continuum. The latter is explained by electron-hole excitations inside a lifetime broadened conduction band. This continuum is directly related to the scattering rates of the charge carriers, which in turn, by a feed back process, lead to the continuum.

\end{abstract}


\maketitle

\section{\label{sec:intro} I. INTRODUCTION}

In a recent paper by Husain {\it et al.}~\cite{Husain2019}, the two-particle electronic excitations in Bi$_2$Sr$_2$CaCu$_2$O$_{8+x}$  have been studied by Electron Energy-Loss Spectroscopy in reflection (R-EELS) in the ``strange metal'' range between underdoped and overdoped materials. The work is a continuation of previous work by the group, which was focused on a more narrow doping range~\cite{Vig2017,Mitrano2018}. The main result of these investigations is that this high-$T_c$ superconductor
does not exhibit well-defined plasmon excitations anywhere in its phase diagram. Rather they have detected a featureless, momentum-independent continuum throughout the strange metal region. They come to the conclusion that a new kind of theory of strongly interacting matter may be needed to explain this continuum, which may also be connected to the phenomena of high-$T_c$ superconductivity. Thus, the ``overdamped'' plasmon was discussed in terms of holographic theories~\cite{Romero-Bermudez2019}, which predicted, different from the classical Landau damping, a strong damping even for the long-wavelength plasmons due to  ``quantum critical fluctuations''. 

The continuum, first detected by Raman spectroscopy~\cite{Bozovic1987,Staufer1990} provides a reasonable explanation for the linear temperature dependence of the resistivity in the normal state, one of the outstanding properties of the strange metals. The reason for this is that the lifetime of the quasi-particles is related to  the energy integral over the constant susceptibility, which leads to a linear energy ($\Omega$) or temperature ($T$) dependence of the scattering rate $\Gamma(\Omega,T)$, or the imaginary part of the complex self-energy 
$\Sigma(\Omega,T)$~\cite{Varma1989}. This is the basis of the conjecture of the marginal Fermi liquid~\cite{Varma1989}. 

Finally, we mention also that the existence of plasmons is interesting because plasmons  may mediate superconductivity as predicted by several theoretical models~\cite{Ruvalds1987,Kresin1988}.

In this Comment, I first review the experimental methods connected with two-particle electronic excitations: EELS in transmission (T-EELS), non-resonant and Resonant Inelastic X-Ray Scattering (IXS, RIXS), optical spectroscopy, and Raman spectroscopy. Then the R-EELS experimental results on cuprates are compared with previous results. Later a model for the collective plasmon  excitations and the single electron-hole ($e-h$) excitations is described, which is based on our  knowledge in this field which was established during the last 50 years: a Lindhard model~\cite{Lindhard1954,Pines1966}, extended by $e-h$ excitations between the conduction band and a band caused by umklapp-scattering from the second into the first Brillouin zone.  A further extension comes from the finite lifetime broadening of the single-particle excitations. Values for this broadening  are taken from Angle-Resolved Photoemission Spectroscopy (ARPES)~\cite{Damascelli2003}. Finally,  this model is compared to experimental results. The conclusion is that the extended Lindhard model describes many details of the experimental results on two-particle excitations in cuprates. 
\begin{figure}[ht!]
\centering
\includegraphics[angle=0,width=1.0\linewidth]{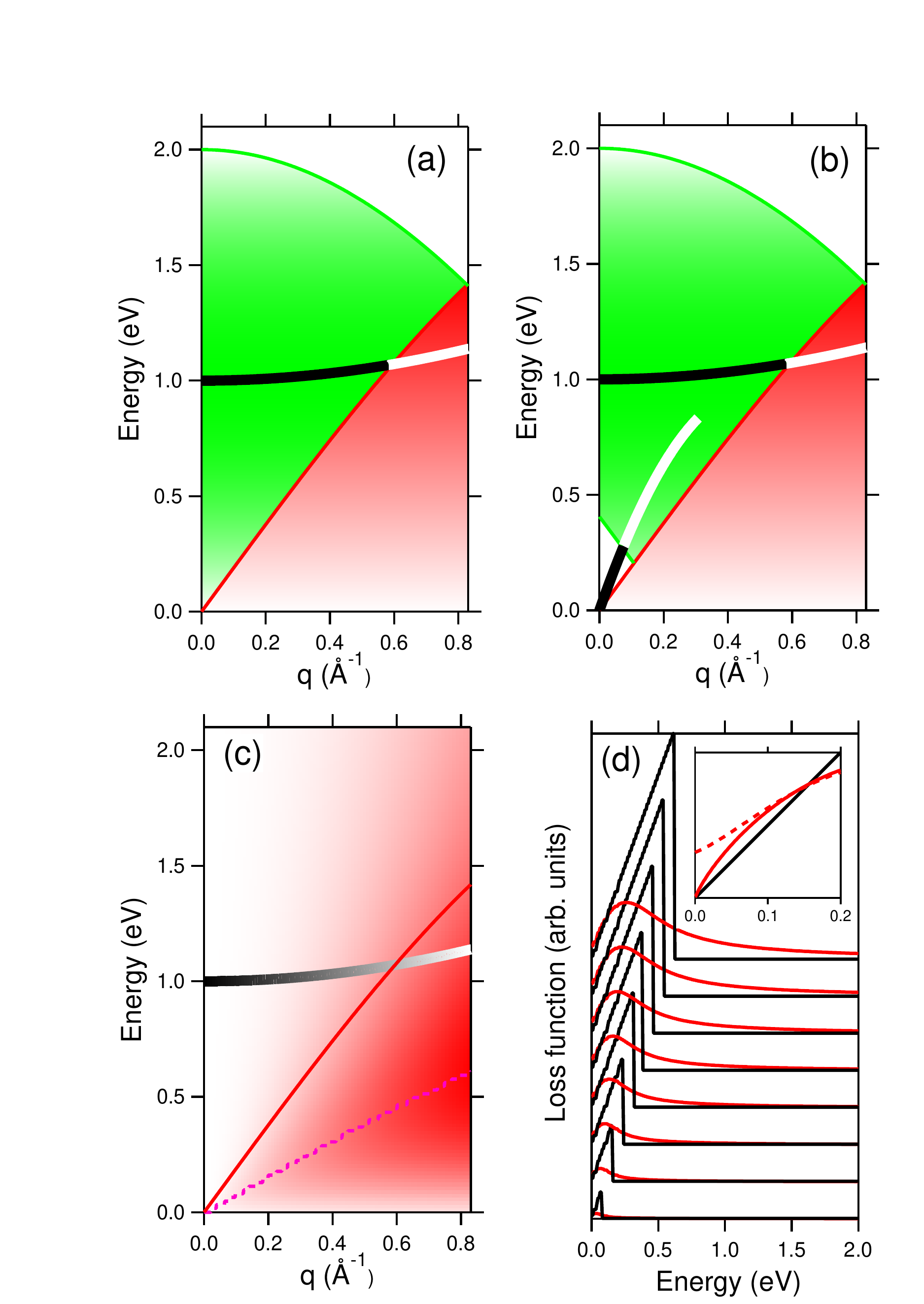}
 \caption{
Sketch of a map of the loss function $\Im(1/\epsilon(\omega,q))$
as a function of energy ($\omega$) and momentum transfer ($q$). Black/white lines: weakly/heavily damped plasmon excitations. Red regions: intra-band electron-hole excitations. Green regions: inter-band transitions due to umklapp scattering. The latter are not shown in the range of intra-band transitions. The size of the loss-function is indicated by the intensity of the green and red color. (a) Extended Lindhard model for intra-, inter-band, and plasmon excitations for a half-filled band without doping. (b) Extended Lindhard model for intra-, inter-band and plasmon excitations for a half-filled band with n-type (or p-type) doping. Acoustic plasmons appear for out-of-plane momentum transfers in the ``empty'' region at low energy and low momentum transfer. (c) Extended Lindhard model taking into account the lifetime broadening of the conduction band. Only spectral weight for the intra-band transitions at zero temperature is shown. Dashed magenta line: maximum of the spectral weight of the continuum of intra-band transitions. (d) Waterfall plots of the energy dependence of the intra-band part of the loss functions for various momentum transfers ranging from $q=0.1$~\AA$^{-1}$ (lowest curve) to $q=0.33~$\AA$^{-1}$ (uppermost curve).
Black curve: Lindhard model. Red curve: extended Lindhard model taking into account the lifetime broadening of the conduction band. Inset in (d): the same as in the main panel but for low energy and $q=0.4~$\AA$^{-1}$. In addition, the inset shows a calculation for $T=300$ K (red dashed line). 
}
\centering
\end{figure}
\section{
II. EXPERIMENTAL METHODS}
Since the 60th of the last century, the most popular method for the direct observation of plasmons was  T-EELS using dedicated spectrometers. It measures the energy ($\omega$) and momomentum ($q$) dependent loss-function $\Im(1/\epsilon(\omega,q))$ which has a maximum at the energy ($\omega_P$) of the plasmon  and the width is related to the lifetime. The method is documented in various review articles~\cite{Raether1965,Daniels1970,Schnatterly1979,Schattschneider1986,Fink1989,Roth2014}. From the very beginning, T-EELS was momentum resolved. The momentum parallel and perpendicular to the incident electron beam is well defined. The best energy resolution is slightly below 0.1 eV. The momentum resolution $\Delta q_{||}$ is about 0.04 \AA$^{-1}$ and   $\Delta q_{\perp}$ is very small for low energy conduction band excitations. According to the Liouville theorem~\cite{Liouville1838}, the high momentum resolution is directly related with the poor spacial resolution. The cross section for inelastic electron scattering is proportional to $1/q^2$. Therefore the method has its strength at low $q$ or  long wavelengths.

T-EELS is very often  performed using Transmission Electron Microscopes (TEM)~\cite{Egerton1996}, which have a high spacial resolution and thus a poor momentum resolution.
At present there are some TEMs which now have a high energy resolution up to 
0.01 eV and a reasonable momentum resolution at the expense of the spacial 
resolution~\cite{Terauchi1999,Krivanek2014,Senga2019,Hong2020}.

The loss-function can be also directly measured by IXS. Also in this case, the cross section is well 
understood~\cite{Schuelke2007}. 
Because it increases with $q^2$, the method is more useful at higher momentum transfer.  The energy and the momentum resolution is up to 0.04 eV and 0.04 \AA$^{-1}$, respectively.

RIXS was applied mostly for the investigation of phononic, magnetic, and non-dipolar electronic excitations~\cite{Kotani2001,Ament2011}.
Recently it was also possible to study acoustic plasmon excitations with an energy resolution of about 0.04 eV~\cite{Hepting2018,Nag2020}.

In the past, R-EELS  was predominantly used for the study of excitations of molecules on the surface of solids~\cite{Ibach1982,Lueth1993}. Recently, the method was also used to study momentum dependent electronic excitations in solids~\cite{Schulte2002,Vig2017}. A very high energy resolution of the order of meV was achieved. On the other hand, the scattering process is rather complicated: the dominant scattering channel is supposed to be an elastic plus inelastic double scattering process~\cite{Mills1975}.
The large momentum perpendicular to the surface (for a primary energy of 50 eV $q_{\perp}\approx 4$ \AA$^{-1}$) may yield a finite  $q_{||}$ for  non-zero energy losses~\cite{Chiarello2000}.

Optical spectroscopy yields the reflectivity or by ellipsometry the complex dielectric function~\cite{Wooten1972}. From those data, using a Kramers-Kronig transformation, the 
loss-function and electron-hole excitations can be derived for zero momentum.

A complementary method is electronic Raman scattering, which  provides anisotropic and zero momentum information for electron-hole excitations~\cite{Devereaux2007}.
\section{
III Comparison of the R-EELS results with other results}
Using EELS in transmission, well defined plasmons close to 1 eV were detected in numerous 2D cuprates: Bi$_2$Sr$_2$CaCu$_2$O$_{8+x}$~\cite{Nuecker1989,Wang1990,Nakai1990,Nuecker1991,Terauchi1999}
, YBa$_2$Cu$_2$O$_7$~\cite{Tarrio1988,Romberg1990}, YBa$_2$Cu$_4$O$_8$~\cite{Knupfer1994}, and 
Ca$_{1.9}$Na$_{0.1}$CuO$_2$Cl$_2$~\cite{Schuster2012}. In compounds containing CuO chains (YBa$_2$Cu$_2$O$_7$ and YBa$_2$Cu$_4$O$_8$), for momentum parallel to the chains,  the plasmon is shifted to higher energies. In all cases, the plasmon was not overdamped. The full width at half maximum (FWHM)  of the plasmon is about 0.5 eV, which is smaller than its energy, signaling a well defined quasi particle. This width is not abnormal -- the plasmon of the nearly free electron metals Na and Al at 6 and 15 eV have a width of 
0.25 eV and 1 eV, respectively~\cite{vom_Felde_1989,Raether1965}. Also in the quasi-1D ladder compounds  (La,Ca)$_x$Sr$_{14-x}$Cu$_{24}$O$_{41}$ well defined plasmons have been detected for $q$ parallel to the ladders~\cite{Roth2010,Roth2020}.

In same cases, the plasmon dispersion has been studied. This dispersion was always positive, typical of an electron liquid with a finite compressibility. This  plasmon dispersion could be well explained in terms of collective intra-band transitions. The anisotropic dispersion provided detailed information on the shape of the Fermi surface~\cite{Nuecker1991,Grigoryan1999,Roth2020}. At higher momentum transfer near $q=0.4$ \AA$^{-1}$ the plasmon is heavily damped and only a continuum is visible. This was explained by Landau damping, i.e., the plasmon decays into intra-band transitions~\cite{Nuecker1989,Wang1990,Nuecker1991}.

On the other hand, EELS using a TEM with an energy resolution of 0.01 eV  showed in cuprates no plasmon, but just a continuum extending to $\approx 1.5$ eV~\cite{Terauchi1999}. 
This result can be explained by the poor momentum resolution in this experiment which leads to a spectral weight predominantly stemming from the high-$q$ loss function, where the plasmon has decayed by Landau damping. 

Early optical spectroscopy on La$_{2-x}$Sr$_x$CuO$_4$ derived for $x$ close to optimal doping a plasmonlike reflectivity edge  near 0.8 eV~\cite{Uchida1991}. More recent results for the loss-function of Bi$_2$Sr$_2$CaCu$_2$O$_{8+x}$, derived from optical data, perfectly agree with the EELS results for momentum transfer close to zero~\cite{Levallois2016}.

Another result which is in stark discrepancy with the recent  R-EELS studies of Ref.~\cite{Husain2019} and which also does not support  theories for heavily damped plasmons due to quantum fluctuations, results from  recent RIXS studies on n- \cite{Hepting2018} and p-type doped cuprates~\cite{Nag2020}. In these experiments acoustic plasmons with an out-of-plane dispersion  have been detected in the low-($\omega,q$) range with a very small width of $\approx 0.1$ eV.

All these findings document that well defined plasmon excitations exist in cuprate based systems. In the following, it is shown that they can be well modeled based upon an extended Lindhard model.

\section{
IV The extended Lindhard model}
In the original Lindhard model~\cite{Lindhard1954}, the collective and the single $e-h$ excitations have been calculated for a free electron system.
For a half-filled band a sketch of the $(\omega,q)$ map for possible charge excitations is presented in Fig.~1(a). The one-dimensional band structure $E(k)=-4tcos(ak)$ was used with $t=0.5$ eV and the lattice constant $a= 3.8$ \AA . This band structure corresponds to that along the nodal direction in cuprates, were the highest Fermi velocity ($v_F$) appears. Because both, the plasmon energy and the single $e-h$ excitation region are related to the highest $v_F$, a two dimensional cuprate band structure would not change very much the map shown in Fig.~1(a). The  $e-h$ intra-band excitations exists in a $2k_F$ wide stripe which is  underlayed by red color.

The collective plasmon excitations appear below a critical momentum $q_c$ showing a dispersion which can be explained in terms of the band structure~\cite{Nuecker1991,Grigoryan1999}. Above $q_c\approx\omega_P/v_F$, the plasmon decays into single $e-h$ excitations, termed Landau damping. 

There should be no plasmon width for long wave-length caused by  correlation effects: the charge of the electrons is just shifted against the charge of the ions and the restoring force is determined by the surface charges. This view is supported by calculations of the plasmon broadening within a free-electron model due to electron-electron interactions~\cite{DuBois_1969}, which derived for the plasmon width $\Gamma_P=Bq^2$. Thus the finite plasmon width at $q=0$ was for many years a puzzle.
 
Finally, it was shown that in non-free-electron systems the plasmon width for long wave lengths is determined by a decay of the plasmons into $e-h$ inter-band transitions~\cite{Paasch1970,Gibbons1977}.  The strength of the inter-band transitions at low momentum transfer  is related to the square of the Fourier transform of the pseudo-potential, which leads to an umklapp scattering of the conduction band from the second to the first Brillouin zone.  This leads  to a separate region of inter-band transitions in the $(\omega,q)$ map, marked in green in Fig.~1(a).

Intra- and inter-band transitions have been detected in several EELS or IXS experiments. Even for  nearly-free electron systems such as alkali metals and Be, both intra- and inter-band transitions are visible by EELS~\cite{vom_Felde_1989} and 
IXS~\cite{Schuelke1987}, 
respectively. In highly correlated systems the dielectric function and thus the loss-function is enhanced by electronic local field effects~\cite{Hubbard1957,Mahan2000}.

Upon doping, the Fermi level is shifted to lower or higher energies. This leads in the $(\omega,q)$ map at low $q$ and low $\omega$ to a triangular ``empty'' region without intra- or inter-band transitions. In Fig.~1(b) we show this case for an electron doped cuprate. As demonstrated for p-type doped graphite~\cite{Ritsko1980} and n-type doped  graphite~\cite{Grunes1983,Roth2013}, very narrow plasmons could be detected in this region. 
 
Another extension of the Lindhard model  is related to the finite lifetime of holes and  electrons in the conduction band. Instead of bands, one should use propagators for the occupied and unoccupied band, which depend on the complex self-energy $\Sigma(\Omega)$~\cite{Varma2017}. Its imaginary part is proportional to the lifetime broadening $\Gamma(\Omega)$.  A poor man's version of the quasiparticle lifetime model~\cite{Green1985} is used here to estimate the influence of the finite lifetime. The energy dependent loss function is convoluted with the energy dependent lifetime broadening $\Gamma(\Omega)$ of the quasi particles. The latter is taken from ARPES derived lifetime broadening of  single particle hole excitations in  the conduction band of optimally doped cuprates~\cite{Valla1999a,Kaminski2001,Koitzsch2004,Reber2019}: $\Gamma(\Omega)\approx \Omega$ typical of a marginal Fermi liquid~\cite{Varma1989}.  This leads to a finite single-particle intra-band susceptibility also in the region of the 
inter-band transitions, which is shown in Fig.~1(c).  Fig.~1(d)  shows a waterfall plot of the spectral weight of the intra-band transition as a function of energies for various momentum transfer. In the Lindhard model, with increasing $q$, there is  an increasing  shark fin like structure. After convolution with finite lifetime effects, there is a strong broadening of the peak, the maximum is shifted to lower energies, and the $q$ dependence of the maximum is strongly reduced (see Fig.~1(c)). For highly correlated systems, even below $q_c$, there should be  a finite plasmon broadening from  a decay into intra-band excitations. Thus the width of the plasmons is determined not only by the decay into inter-band transition, the strength of which are determined by the pseudo-potential, but also by a broadening of the intra-band transitions due to the finite lifetime of the quasi particles of the conduction band.
\section{
V Discussion}
How can we explain the differences between the R-EELS and all the T-EELS 
and optical data. The R-EELS at small momentum transfer are very similar to the T-EELS data at higher $q$. As mentioned above, in R-EELS there is a very large $q_{\perp}$. As long as it is perpendicular to the surface (specular reflection) of a 2D electronic system, it has no influence on the measured loss-function with a finite $q_{||}$. Normally, the samples are aligned into the specular geometry by elastic Bragg reflections, i.e., without energy-loss. However, for a finite energy-loss, the absolute momentum values of  the incident and scattered electrons  differ, which leads to an off-specular scattering process and thus to an additional $q_{||}$~\cite{Chiarello2000}. The enormous sensitivity of the data on the exact specular position was already discussed in   Ref.~\cite{Schulte2002}.  In this way, it is possible to explain the differences  between the T-EELS and the R-EELS data: the latter do not reach  low momentum transfers in case of finite energy losses. This may also explain, why there are only a few R-EELS studies of plasmons and surface plasmons  at higher energy~\cite{Chiarello2000}.

Next, we point out that the more traditional model, discussed here, describes many material specific experimental results. It describes the plasmon energy and its positive and anisotropic dispersion on the basis of collective intra-band transitions~\cite{Nuecker1991,Grigoryan1999}. It can explain the finite width of the plasmon at low $q$ in terms of a decay into inter-band transitions. The model gives a reasonable  $q_c$ value which is connected to the Fermi velocity. The narrow plasmon width of the acoustic plasmons~\cite{Hepting2018,Nag2020} is understandable by the ``empty'' region at low $\omega$ and low $q_{||}$. 

Furthermore, the continuum and its energy dependence can be explained by intra-band transitions (see the red lines in Fig.~1(d)).  At low $q$, which is probably not reached in the R-EELS experiment, the electron-hole excitations exhibit a strong $q$-dependence. On the other hand, at higher $q$, the momentum dependence is strongly reduced (see Fig.~1(c) and (d)). Finally, the strong reduction of the continuum at low energies, observed in Fig.~1 of Ref.~\cite{Husain2019}  when going from optimally doped compounds with a marginal Fermi liquid ($\Gamma\propto\omega$) to overdoped cuprates ($\Gamma\propto\omega^2$) can be explained in the latter by the reduced lifetime broadening at low energies~\cite{Fink2016}. 

Moreover, the strong temperature dependence for optimally and overdoped compounds at low energies (see Fig.~1 of Ref.~\cite{Husain2019}) is related to the fact, that  there the broadening due to the thermal energy is much larger than the broadening due to a finite energy (see inset in Fig.~1(d)).

Summarizing the discussion, the extended Lindhard model together with the experimental results on the lifetime broadening of single particle excitations from ARPES can explain the heavily discussed continuum by intra-band transitions. In turn the continuum in cuprates is related by a feed-back process to  the large scattering rates caused by on-site Coulomb interactions. Thus both the continuum and the finite lifetime broadening are related to the resistivity in the ``strange metal'' range of cuprates.
\section{
VI Conclusion}
Hopefully, this comment will  motivate experimental groups to construct EELS spectrometers with  energy resolutions in the range of some meVs. Spectrometers with a resolution of 4 meV were already built in the 70th of the last century~\cite{Schroeder1972}. As mentioned above, there are also some more recent attempts to achieve such resolutions. Possibly also R-EELS can improve the kinematics of the scattering process to reach data at low momentum transfer. Then it will be possible to bridge the gap with Raman data or compete with RIXS. Moreover, to complement the single-particle excitations from ARPES with two particle excitations from EELS will be an interesting task in future solid state physics.

\section{ ACKNOWLEDGMENTS}
I thank   Lara Benfatto, Bernd B\"uchner, Luis Craco, Stefan-Ludwig Drechsler, Rudi Hackl,  Martin Knupfer, Thomas Pichler, and Friedrich Roth for helpful discussions.

\bibliographystyle{apsrev4-2}
\raggedright
\bibliography{EELS}

\end{document}